\documentclass[aps,preprint]{revtex4}
\usepackage{amsmath}
\usepackage[english]{babel}
\usepackage{amssymb}
\usepackage{graphicx}
\usepackage{epsfig}
\usepackage{bm}
\usepackage{dcolumn}
\usepackage{color}

\def\trmin{{\rm tr}}
\def\mf{{\mbox{\tiny MF}}}
\DeclareMathOperator*{\sumint}{%
\mathchoice%
  {\ooalign{$\displaystyle\sum$\cr\hidewidth$\displaystyle\int$\hidewidth\cr}}
  {\ooalign{\raisebox{.14\height}{\scalebox{.7}{$\textstyle\sum$}}\cr\hidewidth$\textstyle\int$\hidewidth\cr}}
  {\ooalign{\raisebox{.2\height}{\scalebox{.6}{$\scriptstyle\sum$}}\cr$\scriptstyle\int$\cr}}
  {\ooalign{\raisebox{.2\height}{\scalebox{.6}{$\scriptstyle\sum$}}\cr$\scriptstyle\int$\cr}}
}

\begin{document}

\vspace*{2cm}

\title{\sc\Large{Charged pion masses under strong magnetic fields in the NJL model}}

\author{M. Coppola$^{a,b}$, D. G\'omez Dumm$^{c}$  and N.N.\ Scoccola$^{a,b,d}$}

\affiliation{$^{a}$ CONICET, Rivadavia 1917, (1033) Buenos Aires, Argentina}
\affiliation{$^{b}$ Physics Department, Comisi\'{o}n Nacional de Energ\'{\i}a At\'{o}mica, }
\affiliation{Av.\ Libertador 8250, (1429) Buenos Aires, Argentina}
\affiliation{$^{c}$ IFLP, CONICET $-$ Departamento de F\'{\i}sica, Fac.\ de Cs.\ Exactas,
Universidad Nacional de La Plata, C.C. 67, (1900) La Plata, Argentina}
\affiliation{$^{d}$ Universidad Favaloro, Sol{\'{\i}}s 453, (1078) Buenos Aires, Argentina}

\begin{abstract}
The behavior of charged pion masses in the presence of a static uniform
magnetic field is studied in the framework of the two-flavor NJL model,
using a magnetic field-independent regularization scheme. Analytical
calculations are carried out employing the Ritus eigenfunction method, which
allows us to properly take into account the presence of Schwinger phases in
the quark propagators. Numerical results are obtained for definite model
parameters, comparing the predictions of the model with present lattice QCD
results.
\end{abstract}


\maketitle

\renewcommand{\thefootnote}{\arabic{footnote}}
\setcounter{footnote}{0}

The study of the behavior of strongly interacting matter under
intense external magnetic fields has gained increasing interest in
the last few
years~\cite{Kharzeev:2012ph,Andersen:2014xxa,Miransky:2015ava},
especially due to its applications to the analysis of relativistic
heavy ion collisions~\cite{HIC} and the description of compact
objects like magnetars~\cite{duncan}. From the theoretical point
of view, addressing this subject requires to deal with quantum
chromodynamics (QCD) in nonperturbative regimes, therefore,
present analyses are based either in the predictions of effective
models or in the results obtained through lattice QCD (LQCD)
calculations. In this work we focus on the effect of an intense
external magnetic field on $\pi$ meson properties. This issue has
been studied in the last years following various theoretical
approaches for low-energy QCD, such as Nambu-Jona-Lasinio
(NJL)-like
models~\cite{Fayazbakhsh:2013cha,Fayazbakhsh:2012vr,Avancini:2015ady,Zhang:2016qrl,Avancini:2016fgq,Mao:2017wmq,GomezDumm:2017jij,Wang:2017vtn,Liu:2018zag},
chiral perturbation theory~\cite{Andersen:2012zc,Agasian:2001ym}
and path integral
Hamiltonians~\cite{Orlovsky:2013wjd,Andreichikov:2016ayj}. In
addition, results for the light meson spectrum under background
magnetic fields have been recently obtained from LQCD
calculations~\cite{Bali:2011qj,Hidaka:2012mz,Luschevskaya:2014lga,
Bali:2015vua,Bali:2017ian,Bali:2017yku}.

In the framework of the NJL model, mesons are usually described as quantum
fluctuations in the random phase approximation
(RPA)~\cite{Vogl:1991qt,Klevansky:1992qe,Hatsuda:1994pi}, that is, they are
introduced via a summation of an infinite number of quark loops. In the
presence of a magnetic field, the calculation of these loops requires some
special care due to the appearance of Schwinger
phases~\cite{Schwinger:1951nm} associated with each quark propagator. For
the neutral pion these phases cancel out, and as a consequence the usual
momentum basis can be used to diagonalize the corresponding polarization
function~\cite{Fayazbakhsh:2013cha,Fayazbakhsh:2012vr,Avancini:2015ady,Avancini:2016fgq,Mao:2017wmq}.
On the other hand, for the charged pions the Schwinger phases do not cancel,
leading to a breakdown of translational invariance that prevents to proceed
as in the neutral case. In this situation, some existing
calculations~\cite{Zhang:2016qrl,Liu:2018zag} just neglect the Schwinger
phases, taking into account only the translational invariant part of the
quark propagator. Very recently~\cite{Wang:2017vtn}, the use of the
derivative expansion approach has been proposed as an improved approximation
to deal with this issue. It should be noticed, however, that such an
approach is expected to be less reliable as the mass of the meson and/or the
magnetic field increase. The aim of the present work is to introduce a
method that allows us to fully take into account the translational-breaking
effects introduced by the Schwinger phases in the calculation of the charged
meson masses in the RPA approach. Our method is based on the Ritus
eigenfunction approach~\cite{Ritus:1978cj} to magnetized relativistic
systems, which, as we show below, allows us to fully diagonalize the charged
pion polarization function.

We start by considering the Euclidean Lagrangian density for the NJL
two-flavor model in the presence of an electromagnetic field. One has
\begin{equation}
{\cal L} \ = \ \bar \psi \left(- i\, \rlap/\!D + m_0 \right) \psi - G \left[
(\bar\psi \, \psi)^2 + (\bar\psi\, i\gamma_{5}\vec{\tau}\,\psi) \right] \ ,
\label{lagrangian}
\end{equation}
where $\psi = (u\ d)^T$, $\tau_i$ are the Pauli matrices, and $m_0$ is the
current quark mass, which is assumed to be equal for $u$ and $d$ quarks. The
interaction between the fermions and the electromagnetic field ${\cal
A}_\mu$ is driven by the covariant derivative
\begin{equation}
D_\mu\ = \ \partial_{\mu}-i\,\hat Q \mathcal{A}_{\mu}\ ,
\label{covdev}
\end{equation}
where $\hat Q=\mbox{diag}(q_u,q_d)$, with $q_u=2e/3$ and $q_d = -e/3$, $e$
being the proton electric charge.

Since we are interested in studying meson properties, it is convenient to
bosonize the fermionic theory, introducing scalar and pseudoscalar fields
$\sigma(x)$ and $\vec{\pi}(x)$ and integrating out the fermion fields. The
bosonized Euclidean action can be written as~\cite{Klevansky:1992qe}
\begin{equation}
S_{\mathrm{bos}}\ = \ -\log\det\mathcal{D}+\frac{1}{4G}
\int d^{4}x\
\Big[\sigma(x)\sigma(x)+ \vec{\pi}(x)\cdot\vec{\pi}(x)\Big]\ ,
\label{sbos}
\end{equation}
with
\begin{equation}
\mathcal{D}_{x,x'} \ = \ \delta^{(4)}(x-x')\,\big[-i\,\rlap/\!D + m_0 +
\sigma(x) + i\,\gamma_5\,\vec{\tau}\cdot\vec{\pi}(x) \big]\ ,
\label{dxx}
\end{equation}
where a direct product to an identity matrix in color space is understood.
We will consider the particular case of an homogenous stationary magnetic
field $\vec B$ along the 3 axis. Then, choosing the Landau gauge, we have
$\mathcal{A}_\mu = B\, x_1\, \delta_{\mu 2}$.

We proceed by expanding the bosonized action in powers of the fluctuations
$\delta\sigma(x)$ and $\delta\pi_i(x)$ around the corresponding mean field
(MF) values. As usual, we assume that the field $\sigma(x)$ has a nontrivial
translational invariant MF value $\bar{\sigma}$, while the vacuum
expectation values of pseudoscalar fields are zero. Thus we write
\begin{equation}
\mathcal{D}_{x,x'} \ = \ \mathcal{D}^{\mbox{\tiny MF}}_{x,x'} + \delta\mathcal{D}_{x,x'}\ .
\label{dxxp}
\end{equation}
The MF piece is flavor diagonal. It can be written as
\begin{equation}
\mathcal{D}^\mf_{x,x'} \ = \ {\rm diag}\big(\mathcal{D}^{\mf,u}_{x,x'}\, ,\,
\mathcal{D}^{\mf,d}_{x,x'}\big)\ ,
\end{equation}
where
\begin{equation}
\mathcal{D}^{\mf,f}_{x,x'} \ = \ \delta^{(4)}(x-x') \left( - i \rlap/\partial
- q_f \, B \, x_1 \, \gamma_2 + m_0 + \bar\sigma \right).
\end{equation}
On the other hand, the second term in the right hand side of
Eq.~(\ref{dxxp}) is given by
\begin{equation}
\delta D_{x,x'} \ =\ \delta^{(4)}(x-x') \,
 \begin{pmatrix}
   \delta\sigma(x) + i\gamma_5 \delta\pi_0(x) & \sqrt{2}i\gamma_5\,\delta\pi^+(x) \\
    \sqrt{2}i\gamma_5\,\delta\pi^-(x) & \delta\sigma(x) - i\gamma_5 \delta\pi_0(x)
  \end{pmatrix}\ ,
  \label{DeltaD}
\end{equation}
where $\pi^\pm=\left(\pi_1 \mp i \pi_2\right)/\sqrt{2}$. Replacing in the
bosonized effective action and expanding in powers of the meson fluctuations
around the MF values, we get
\begin{eqnarray}
S_{\mathrm{bos}} \ = \ S^{\mbox{\tiny MF}}_{\mathrm{bos}} \, + \,
S^{\mbox{\tiny quad}}_{\mathrm{bos}}\, + \,\dots
\end{eqnarray}
Here, the mean field action per unit volume reads
\begin{equation}
\frac{S^{\mbox{\tiny MF}}_{\mathrm{bos}}}{V^{(4)}} \ = \ \frac{ \bar
\sigma^2}{4 G} - \frac{N_c}{V^{(4)}} \sum_{f=u,d} \int d^4x \, d^4x' \
\trmin\, \ln \left(\mathcal{S}^{\mbox{\tiny MF},f}_{x,x'}\right)^{-1} \ ,
\label{seff}
\end{equation}
where $\trmin$ stands for the trace in Dirac space. The quadratic
contribution is given by
\begin{equation}
S^{\mbox{\tiny quad}}_{\mathrm{bos}}\ = \ \dfrac{1}{2}
\sum_{M=\sigma,\pi^0,\pi^\pm} \int d^4x \, d^4x'\ \delta M
(x)^\ast \left[  \frac{1}{2 G}\; \delta^{(4)}(x-x') - J_M(x,x')
\right] \delta M(x')\ , \label{actionquad}
\end{equation}
where
\begin{eqnarray}
J_{\pi^0} (x,x') &=& N_c \sum_f \trmin \bigg[
\mathcal{S}^{\mf,f}_{x,x'} \ \gamma_5 \ \mathcal{S}^{\mf,f}_{x',x}
\ \gamma_5 \ \bigg]\ ,
\nonumber \\
J_{\pi^\pm} (x,x') &=& 2 N_c  \, \trmin \bigg[
\mathcal{S}^{\mf,u}_{x,x'} \ \gamma_5 \ \mathcal{S}^{\mf,d}_{x',x}
\ \gamma_5 \ \bigg]\ , \label{jotas}
\end{eqnarray}
while the expression for $J_\sigma$ is obtained from that of
$J_{\pi^0}$ just replacing $\gamma_5$ with the unit matrix in Dirac
space. In these expressions we have introduced the mean field
quark propagators $\mathcal{S}^{\mf,f}_{x,x'} = \big(
\mathcal{D}^{\mf,f}_{x,x'} \big)^{-1}$. As is well known, their
explicit form can be written in different
ways~\cite{Andersen:2014xxa,Miransky:2015ava}. For convenience we
take the form in which $\mathcal{S}^{\mf,f}_{x,x'}$ is given by a
product of a phase factor and a translational invariant function,
namely
\begin{equation}
S^{\mf,f}_{x,x'} \ = \ e^{i\Phi_f(x,x')}\,\int_p e^{i\, p\, (x-x')}\, \tilde S_p^f\
,
\label{sfx}
\end{equation}
where $\Phi_f(x,x')=\exp\big[i q_f B (x_1+x_1')(x_2-x_2')/2 \big]$ is the
so-called Schwinger phase. We have introduced here the shorthand notation
\begin{equation}
\int_{p}\ \equiv \ \int \dfrac{d^4 p}{(2\pi)^4}\ .
\label{notation1}
\end{equation}

We express now $\tilde S_p^f$ in the Schwinger form
\cite{Andersen:2014xxa,Miransky:2015ava}
\begin{eqnarray}
\tilde S_p^f & = & \int_0^\infty d\tau\
\exp\!\bigg[-\tau\Big(M^2+p_\parallel^2+p_\perp^2\,\dfrac{\tanh\tau B_f}{\tau B_f}\Big) \bigg] \times \nonumber\\
& & \bigg\{\big(M-p_\parallel \cdot \gamma_\parallel\big)\,\big[1+i s_f \ \gamma_1 \gamma_2 \ \tanh\tau B_f\big] -
\dfrac{p_\perp \cdot \gamma_\perp}{\cosh^2 \tau B_f}  \bigg\}\ .
\label{sfp_schw}
\end{eqnarray}
Here we have used the following definitions. The perpendicular and parallel
gamma matrices are collected in vectors $\gamma_\perp = (\gamma_1,\gamma_2)$
and $\gamma_\parallel = (\gamma_3,\gamma_4)$, and, similarly, we have
defined $p_\perp = (p_1,p_2)$ and $p_\parallel = (p_3,p_4)$. The quark
effective mass $M$ is given by $M=m_0+\bar\sigma$, and other definitions are
$s_f = {\rm sign} (q_f B)$ and $B_f=|q_fB|$. Notice that the integral in
Eq.~(\ref{sfp_schw}) is divergent and has to be properly regularized, as we
discuss below.

Replacing the above expression for the quark propagator in Eq.~(\ref{seff})
and minimizing with respect to $M$ we obtain the gap
equation~\cite{Klevansky:1989vi}
\begin{equation}
M \ = \ m_0 + 4 G M N_c\, I\ ,
\label{gapeq}
\end{equation}
where
\begin{eqnarray}
I \ = \ \frac{1}{8 \pi^2} \sum_f  \int_0^\infty \frac{d\tau}{\tau^2} \ e^{-\tau M^2}  \tau B_f \coth( \tau B_f
)\ .
\end{eqnarray}
To regularize the above integral we use here the Magnetic Field Independent
Regulation (MFIR) scheme~\cite{Menezes:2008qt,Allen:2015paa}. That is, we
subtract from $I$ the unregulated integral in the $B=0$ limit, $I_{B=0}$,
and then we add it in a regulated form $I^{\rm (reg)}_{B=0}$. Thus, we have
\begin{equation}
I^{\rm (reg)} \ = \ I^{\rm (reg)}_{B=0} + I^{\rm (mag)}\
,
\end{equation}
where $I^{\rm (mag)}$ is a finite, magnetic field dependent contribution given by
\begin{eqnarray}
I^{\rm (mag)} & = &  \frac{1}{8 \pi^2} \sum_f  \int_0^\infty
\frac{d\tau}{\tau^2} \ e^{-\tau M^2} \left[  \tau B_f \coth( \tau B_f ) -1
\right]\nonumber \\
& = & \frac{M^2}{8\pi^2} \sum_f \left[ \frac{ \ln \Gamma(x_f)}{x_f} -
\frac{\ln 2\pi}{2x_f} + 1 - \left( 1 - \frac{1}{2x_f}\right) \ln x_f
\right]\ ,
\label{imag}
\end{eqnarray}
with $x_f = M^2/(2 B_f)$. This expression is in agreement with the
corresponding one given in Ref.~\cite{Klevansky:1992qe}, and it also matches
the result obtained in Ref.~\cite{Menezes:2008qt}, where the propagator is
expressed in terms of a sum over Landau levels. On the other hand, the
regulated piece $I^{\rm (reg)}_{B=0}$ does depend on the regularization
prescription. Choosing the standard procedure in which one introduces a 3D
momentum cutoff $\Lambda$, we get the well known result
\cite{Klevansky:1992qe}
\begin{equation}
I^{\rm (reg)}_{B=0} \ = \ I_1 \equiv \ \frac{1}{2 \pi^2} \left[
\sqrt{ \Lambda^2 + M^2 } + M^2\;
\ln{\left( \frac{ M}{\Lambda + \sqrt{\Lambda^2 + M^2}}\right)}\right]\ . \label{I13d}
\end{equation}

\hfill

Let us turn now to the determination of pion masses, starting by
the simpler case of the neutral pion $\pi^0$. We notice that the
analysis of the $\pi^0$ pole mass in the presence of a magnetic
field within the MFIR scheme has already been carried out in
Refs.~\cite{Avancini:2015ady,Avancini:2016fgq}. However, in those
works the authors use a representation of the quark propagator
different from the Schwinger one in
Eqs.~(\ref{sfx}-\ref{sfp_schw}). Thus, we find it opportune to
verify that both representations lead to the same results for the
$\pi^0$ mass. The study of the $\sigma$ sigma meson mass can be
performed in an entirely equivalent way, and will not be
considered here. We start by replacing Eq.~(\ref{sfx}) in the
expression for the polarization function $J_{\pi^0}(x,x')$ in
Eq.~(\ref{jotas}). This leads to
\begin{equation}
J_{\pi^0}(x,x') \ = \ N_c \sum_f \int_{pp'} \trmin\, (\tilde S_p^f \,
\gamma_5 \, \tilde S_{p'}^f \, \gamma_5)\ e^{i\Phi_f(x,x')}\
e^{i\Phi_f(x',x)}\ e^{i (p-p')(x-x')} \  .
\label{J0x}
\end{equation}
Notice that the contributions of Schwinger phases to each term of the sum
correspond to the same quark flavor, hence, they cancel out. As a
consequence, the polarization function depends only on the difference $x-x'$
(i.e., it is translational invariant), which leads to the conservation of
$\pi^0$ momentum. If we take now the Fourier transform of the $\pi^0$ fields
to the momentum basis, the corresponding transform of the polarization
function will be diagonal in $q,q'$ momentum space. Thus, the $\pi^0$
contribution to the quadratic action in the momentum basis can be written as
\begin{equation}
S^{\mbox{\tiny quad}}_{\pi^0} \ = \ \dfrac{1}{2} \int_{q} \
\delta\pi^0(-q) \, \left[\frac{1}{2 G} -
J_{\pi^0}(q_\perp^2,q_\parallel^2) \right] \delta\pi^0(q)\ ,
\label{actionquadpi0p}
\end{equation}
where
\begin{equation}
J_{\pi^0}(q_\perp^2,q_\parallel^2) \ = \ N_c \sum_f \int_{p}
\trmin\, (\tilde S_{p_+}^f \, \gamma_5 \, \tilde S_{p_-}^f \,
\gamma_5)\ , \label{J0q}
\end{equation}
with $p_\pm = p \pm q/2$. Choosing the frame in which the $\pi^0$ meson is
at rest, its mass can be obtained by solving the equation
\begin{equation}
\frac{1}{2 G} - J_{\pi^0}(0,-m_{\pi^0}^2) \ = \ 0\ .
\end{equation}
Replacing Eq.~(\ref{sfp_schw}) into Eq.~(\ref{J0q}), a straightforward calculation leads to
\begin{eqnarray}
J_{\pi^0}(q_\perp^2,q_\parallel^2) &=& \dfrac{N_c}{4\pi^2} \sum_f\ B_f \int_0^\infty dz \int_{0}^1 dy \ e^{-z\left[M^2+y(1-y)q_\parallel^2 \right]}\,\times  \nonumber \\
& & \exp\left[-\dfrac{q_\perp^2}{B_f} \dfrac{\sinh(yzB_f)\sinh[(1-y)zB_f]}{\sinh(zB_f)} \right]
\times\nonumber \\
& & \Bigg\{ \left[M^2+\dfrac{1}{z}-y(1-y)q_\parallel^2 \right] \coth(zB_f) \, + \nonumber \\
& & \dfrac{B_f}{\sinh^2(zB_f)} \left[1 - \dfrac{q_\perp^2}{B_f} \dfrac{\sinh(yzB_f)\sinh[(1-y)zB_f]}{\sinh(zB_f)} \right] \Bigg\}\ .
\label{J0B}
\end{eqnarray}
This expression can be also derived from Eq.~(2.14) of
Ref.~\cite{Klevansky:1991ey}. As usual, here we have used the changes of
variables $\tau = y z$ and $\tau'=(1-y)z$, $\tau$ and $\tau'$ being the
integration parameters associated with the quark propagators. As done at the
MF level, we regularize the above integral using the MFIR scheme. That is,
we subtract the corresponding unregulated contribution in the $B=0$ limit,
given by
\begin{equation}
J_{\pi,B=0} (q^2) \ = \ \dfrac{N_c}{2\pi^2} \int_0^\infty
\dfrac{dz}{z} \int_{0}^1 dy \ e^{-z\left[M^2+y(1-y)q^2 \right]}
\left[M^2+\dfrac{2}{z}-y(1-y)q^2 \right]\ , \label{TPFB0}
\end{equation}
and add it in a regularized form $J_{\pi,B=0}^{\rm (reg)} (q^2)$. The
regularized polarization function is then given by
\begin{equation}
J^{\rm (reg)}_{\pi^0}(q_\perp^2,q_\parallel^2) \ = \ J^{\rm (reg)}_{\pi,B=0}
(q^2) + J^{\rm (mag)}_{\pi^0}(q_\perp^2,q_\parallel^2)\ .
\end{equation}
{}From Eqs.~(\ref{J0B}) and (\ref{TPFB0}), the finite magnetic
field-dependent term $J^{\rm (mag)}_{\pi^0}(q_\perp^2,q_\parallel^2)$,
evaluated at $q_\perp^2=0$, $q_\parallel^2 = -m_{\pi^0}^2$, is easily found
to be
\begin{eqnarray}
J^{\rm (mag)}_{\pi^0}(0,-m_{\pi^0}^2)  &=& \dfrac{N_c}{4\pi^2}
\sum_f B_f \int_0^\infty dz \int_{0}^1 dy
\ e^{-z\left[M^2-y(1-y)m_{\pi^0}^2\right]} \nonumber \\
&& \hspace{-2cm}
\times\; \Bigg\{\left[M^2+\dfrac{1}{z}+y(1-y)m_{\pi^0}^2 \right] \left[ \dfrac{B_f}{\tanh(zB_f)} - \dfrac{1}{z} \right] + \dfrac{B_f}{\sinh^2(zB_f)} - \dfrac{1}{z^2} \Bigg\}\ .
\label{J0magqp2}
\end{eqnarray}
On the other hand, to get $J^{\rm (reg)}_{\pi,B=0} (q^2)$ we can use the 3D
momentum cutoff scheme, as done in the case of the gap equation. One has in
this way
\begin{equation}
J^{\rm (reg)}_{\pi,B=0}(q^2) \ = \ 2 N_c \left[ I_1 + q^2 I_2(q^2) \right]\ ,
\label{J+0reg}
\end{equation}
where $I_1$ is given by Eq.~(\ref{I13d}), while
\begin{eqnarray}
\hspace{-0.5cm} I_2(q^2) &=& \frac{1}{4 \pi^2} \int_0^1 dy \left[
\frac{\Lambda}{\sqrt{\Lambda^2+ M^2 + y(1-y) q^2}}\, +\, \ln{ \frac{
{\sqrt{M^2 + y(1-y) q^2}}}{\Lambda + \sqrt{\Lambda^2 +
M^2 + y(1-y) q^2}}} \right] .
\label{I23d}
\end{eqnarray}

It is interesting to note that, after some changes of variables (and making
use of the gap equation), our result for $J^{\rm
(reg)}_{\pi^0}(0,-m_{\pi^0}^2)$ is shown to be in agreement with the
corresponding expression obtained in Ref.~\cite{Avancini:2015ady}, where the
calculation has been done using an expansion in Landau levels for the quark
propagators (instead of considering the Schwinger form in
Eq.~(\ref{sfp_schw})). Since both calculations use the 3D cutoff
regularization for the $B = 0$ piece, it is seen that different
representations of the quark propagator lead to the same result for the
(finite) magnetic dependent piece $J^{\rm (mag)}_{\pi^0}(0,-m_{\pi^0}^2)$,
as they should.

\hfill

Finally, we discuss how to treat the case of the charged pions, which is, in
fact, the main topic of this work. For definiteness we consider the
$\pi^+$ meson, although a similar analysis, leading to the same expression
for the $B$-dependent mass, can be carried out for the $\pi^-$. As in the
case of the $\pi^0$, we start by replacing Eq.~(\ref{sfx}) in the expression
of the corresponding polarization function in Eq.~(\ref{jotas}). We get
\begin{eqnarray}
J_{\pi^+}(x,x') \ = \ 2N_c \int_{pp'} \trmin_D (\tilde S_p^u \, \gamma_5 \, \tilde S_{p'}^d \, \gamma_5)\,
e^{i\Phi_u(x,x')}\, e^{i\Phi_d(x',x)}\, e^{i (p-p')(x-x')}\  .
\label{J+q1}
\end{eqnarray}
Contrary to the neutral case, here the Schwinger phases do not cancel, due
to their different quark flavors. Therefore, this polarization function is
not translational invariant, and consequently it will not become diagonal
when transformed to the momentum basis. In this situation we find it
convenient to follow the Ritus eigenfunction method~\cite{Ritus:1978cj}.
Namely, we expand the charged pion field as
\begin{equation}
\pi^+(x) \ = \ \sumint_{\bar q}\ \mathbb{F}_{\bar q}^+(x) \,
\pi_{\bar q}^+ \ ,
\label{Ritus}
\end{equation}
where we have used the shorthand notation
\begin{equation}
\sumint_{\bar q}\ \equiv \ \dfrac{1}{2\pi}\sum_{k=0}^\infty \int_{q_2q_3q_4}
\ ,\qquad \bar q \equiv (k,q_2,q_3,q_4) \ ,
\end{equation}
and the functions $\mathbb{F}_{\bar q}^+(x)$ are given by
\begin{equation}
\mathbb{F}_{\bar q}^+(x) \ = \ N_k \, e^{i ( q_2 x_2 + q_3 x_3 + q_4 x_4)}
\, D_k(\rho_{+})\ .
\label{Fq}
\end{equation}
Here $D_k(x)$ are the cylindrical parabolic functions, and we have used the
definitions $N_k= (4\pi B_{\pi^+})^{1/4}/\sqrt{k!}$ and $\rho_+ =
\sqrt{2B_{\pi^+}}\,x_1-s_+\sqrt{2/B_{\pi^+}}\,q_2$, where $B_{\pi^+} =
|q_{\pi^+} B|$ and $s_+= \mathrm{sign}(q_{\pi^+} B)$, with $q_{\pi^+} = q_u
- q_d = e$. A rather long but straightforward calculation shows that in this
basis the charged pion polarization function is diagonal. We find that the
corresponding contribution to the quadratic action in Eq.~(\ref{actionquad})
is given by
\begin{equation}
S^{\mbox{\tiny quad}}_{\pi^+} \ = \ \dfrac{1}{2} \sumint_{\bar q}
\ (\delta\pi_{\bar q}^+)^\ast \left[ \frac{1}{2G} -
J_{\pi^+}(k,\Pi^2)\right] \delta\pi_{\bar q}^+ \ ,
\label{actionquadTPF}
\end{equation}
where
\begin{eqnarray}
J_{\pi^+}(k,\Pi^2) &=& \dfrac{N_c}{2\pi^2} \int_0^\infty\! dz
\int_{0}^1 dy \ \frac{1}{\alpha_+}\, \,
e^{-zM^2-zy(1-y)\left[\Pi^2-(2k+1)\ B_{\pi^+} \right]}
\left(\dfrac{\alpha_-}{\alpha_+}\right)^k \times
\nonumber \\
&& \Bigg\{\left[M^2+\dfrac{1}{z}-y(1-y)\left(\Pi^2-(2k+1)\,
B_{\pi^+} \right) \right](1-t_u \,t_d) \,+
\nonumber \\
&& \dfrac{(1-t_u^2)(1-t_d^2)}{\alpha_+\,\alpha_-}\, \Big[ \alpha_- + (\alpha_- - \alpha_+)\,k \Big] \Bigg\}\ .
\label{J+B}
\end{eqnarray}
Here we have introduced the definitions $\Pi^2=(2k+1)\, B_{\pi^+}
+q_\parallel^2$, $t_u=\tanh (B_u y z)$, $t_d=\tanh [B_d (1-y) z]$ and
$\alpha_\pm = (B_d t_u+B_u t_d \pm
 B_{\pi^+} \,t_ut_d)/(B_u B_d)$.

As in the case of the neutral pion, the polarization function in
Eq.~(\ref{J+B}) turns out to be divergent and has to be regularized. Once
again, this can be done within the MFIR scheme. However, due to quantization
in the 1-2 plane this requires some care, viz.~the subtraction of the $B=0$
contribution to the polarization function has to be carried out once the
latter has been written in terms of the squared canonical momentum $\Pi^2$,
as in Eq.~(\ref{J+B}). Thus, the regularized $\pi^+$ polarization function
is given by
\begin{equation}
J_{\pi^+}^{{\rm (reg)}}(k,\Pi^2) \ = \ \, J^{\rm
(reg)}_{\pi,B=0}(\Pi^2) \, +  \, J_{\pi^+}^{{\rm (mag)}}(k,\Pi^2)\
, \label{J+reg}
\end{equation}
where
\begin{eqnarray}
J_{\pi^+}^{{\rm (mag)}}(k,\Pi^2) &=& \dfrac{N_c}{2\pi^2}
\int_0^\infty\! dz \int_{0}^1 dy \ e^{-z\,[M^2+y(1-y)\Pi^2]}
\,\times
\nonumber \\
&& \Bigg\{ \bigg[ M^2 +\dfrac{1}{z} - y(1-y)\Big[\Pi^2-(2k+1)\,B_{\pi^+}\Big] \bigg] \times \nonumber \\
&& \bigg[\dfrac{(1-t_u \,t_d)}{\alpha_+}\,
\bigg(\dfrac{\alpha_-}{\alpha_+} \bigg)^k\!
e^{z\,y(1-y)(2k+1)B_{\pi^+}} - \dfrac{1}{z} \bigg] +
\nonumber \\
&& \dfrac{(1-t_u^2)(1-t_d^2)}{\alpha_+^2\,\alpha_-}\, \bigg(\dfrac{\alpha_-}{\alpha_+} \bigg)^k
\Big[ \alpha_- + (\alpha_- - \alpha_+)\,k \Big]\;e^{z\,y(1-y)(2k+1)\,B_{\pi^+}} - \nonumber \\
&& \dfrac{1}{z} \bigg[ \dfrac{1}{z} - y(1-y)(2k+1)\,B_{\pi^+}
\bigg] \Bigg\}\ . \label{J+mag}
\end{eqnarray}
It is easy to see that the integral is convergent in the limit $z\to 0$. The
same expression for $J_{\pi^+}^{{\rm (mag)}}(k,\Pi^2)$ should be obtained if
the propagators are expressed in terms of a sum over Landau levels, although
analytical calculations could be much more cumbersome. On the other hand,
the expression for the subtracted $B=0$ piece is the same as in the $\pi^0$
case, Eq.~(\ref{TPFB0}), replacing $q^2\to \Pi^2$. Therefore, using 3D
cutoff regularization, the function $J^{\rm (reg)}_{\pi,B=0}$ in
Eq.~(\ref{J+reg}) will be given by Eq.~(\ref{J+0reg}). It can be easily seen
that the same polarization function is obtained for the case of the $\pi^-$
meson.

Given the regularized polarization function, we can now derive an equation
for the $\pi^+$ meson pole mass in the presence of the magnetic field. To do
this, let us firstly consider a point-like pion. For such a particle, in
Euclidean space, the two-point function will vanish (i.e., the propagator
will have a pole) when
\begin{equation}
\Pi^2 = - m_{\pi^+}^2\ ,
\label{meB}
\end{equation}
or, equivalently, $q_\parallel^2 = - [ m_{\pi^+}^2 + (2k+1)\,eB]$, for a
given value of $k$. Therefore, in our framework the
charged pion pole mass can be obtained for each Landau level $k$ by solving the
equation
\begin{equation}
\frac{1}{2G} - J_{\pi^+}^{{\rm (reg)}}(k,-m_{\pi^+}^2) \ = \ 0 \ .
\end{equation}
Of course, while for a point-like pion $m_{\pi^+}$ is a
B-independent quantity (the $\pi^+$ mass in vacuum), in the
present model ---which takes into account the internal quark
structure of the pion--- this pole mass turns out to depend on the
magnetic field. Instead of dealing with this quantity, it has
become customary in the literature to define the $\pi^+$
``magnetic field-dependent mass'' as the lowest
quantum-mechanically allowed energy of the $\pi^+$ meson, namely
\begin{equation}
E_{\pi^+}(eB) \ = \  \sqrt{m_{\pi^+}^2 + (2k+1)\,eB + q_3^2}\;\Big|_{q_3 =0,\,k=0}
\ = \ \sqrt{m_{\pi^+}^2 + eB} \
\label{epimas}
\end{equation}
(see e.g.~Ref.~\cite{Bali:2017ian}). Notice that this ``mass'' is magnetic
field-dependent even for a point-like particle. In fact, owing to zero-point
motion in the 1-2 plane, even for $k=0$ the charged pion cannot be at rest
in the presence of the magnetic field.

To get numerical predictions for the behavior of pion masses in the presence
of the $\vec B$ field it is necessary to take a definite parameterization of
the NJL model. In this sense, in addition to usual requirements for the
description of low-energy phenomenological properties, we find it adequate
to choose a parameter set that takes into account LQCD results for the
behavior of quark-antiquark condensates under an external magnetic field. It
is easy to see that at the MF level the quark-antiquark condensates are
given by
\begin{equation}
\langle \bar f f\rangle_B \ = \ -\, N_c\; \trmin \int d^4 x\
\mathcal{S}^{\mf,f}_{x,x} \ = \ -\,\frac{N_c M}{4 \pi^2} \int_0^\infty
\frac{d\tau}{\tau^2} \ e^{-\tau M^2}  \tau B_f \coth( \tau B_f)\ .
\end{equation}
This integral can be easily regulated within the MFIR scheme, as
discussed above. In order to compare with LQCD results given in
Refs.~\cite{Bali:2012zg} we introduce the quantities
\begin{equation}
\Delta \bar\Sigma(B) \ \equiv \ \dfrac{\Delta \Sigma_u (B)+\Delta \Sigma_d (B)}{2}
\ , \qquad
\Sigma^-(B) \ = \ \Delta \Sigma_u (B)- \Delta \Sigma_d (B)\ ,
\end{equation}
where $\Delta\Sigma_f(B) =  - 2\,m_0
\Big[\langle \bar f f \rangle_B - \langle \bar f f \rangle_0 \Big]/D^4$.
Here $D=(135\times 86)^{1/2}$~MeV is a phenomenological normalization constant.

\begin{figure}[htb]
\centering{}\includegraphics[width=0.7\textwidth]{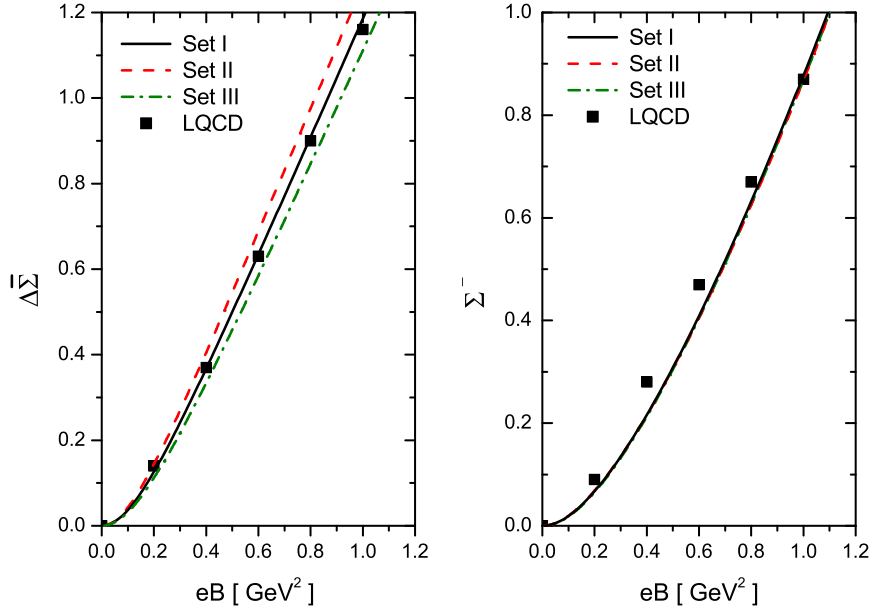}
\caption{(Color online) Left and right panels show the behavior of
$\Delta \bar\Sigma$ and $\Sigma^-$, respectively, as functions of
$eB$ for three different model parameter sets. Results from
lattice QCD calculations~\cite{Bali:2012zg} are included for
comparison.} \label{fig1}
\end{figure}

Let us consider the parameter set $m_0 = 5.66$ MeV, $\Lambda = 613.4$ MeV
and $G\Lambda^2 = 2.250$, which (for vanishing external field) corresponds
to an effective mass $M=350$~MeV and a quark-antiquark condensate $\langle
\bar f f\rangle_0 = (-243.3\ {\rm MeV})^3$. This parameterization, which we
denote as Set I, is shown to properly reproduce the empirical values of the
pion mass and decay constant in vacuum, namely $m_\pi=138$~MeV and
$f_\pi=92.4$~MeV. It also provides a very good agreement with lattice
calculations in Ref.~\cite{Bali:2012zg} for the normalized average
condensate $\Delta \bar\Sigma(B)$. This is shown in the left panel of
Fig.~\ref{fig1}, where the solid line and the fat squares correspond to the
predictions for Set I and LQCD results, respectively. To test the
sensitivity of our results with respect to the model parametrization we have
also considered two alternative parameterizations, denoted as Set II and Set
III, which correspond to $M=320$ and 380~MeV, respectively. In the right
panel of the figure we plot our results for $\Sigma^-(B)$, which also appear
to be consistent with LQCD results~\cite{Bali:2012zg}. It is also seen that
our predictions are not significantly affected by the parameter choice.

\begin{figure}[htb]
\centering{}\includegraphics[width=0.7\textwidth]{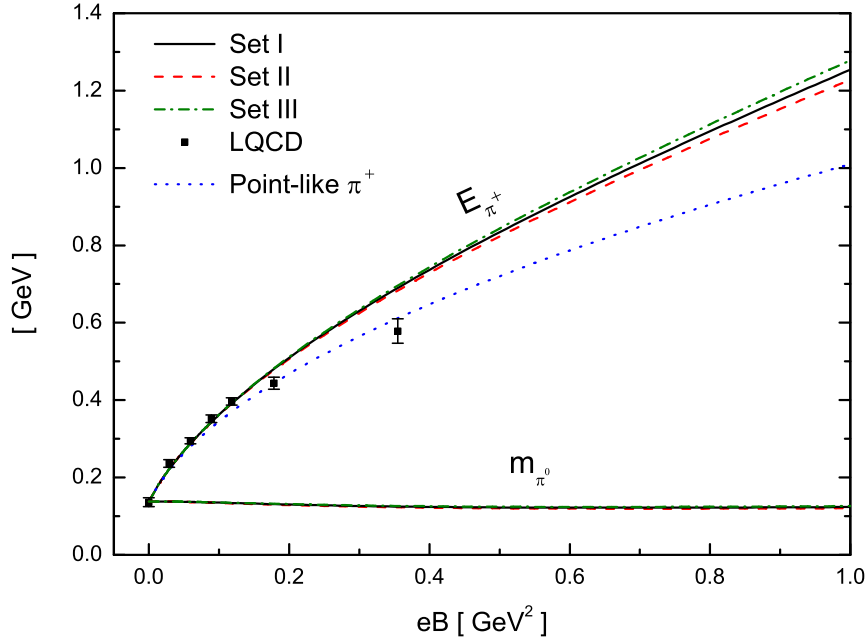} \caption{(Color
online) Neutral pion mass and magnetic field-dependent charged pion mass as
functions of $eB$ for three different model parameter sets (notice that they
are practically indistinguishable from each other in the case of the neutral
pion). For comparison, the behavior of the magnetic field-dependent mass of
a point-like charged pion (dotted line), as well as results from lattice QCD
calculations in Ref.~\cite{Bali:2011qj} (squares) are also included.}
\label{fig2}
\end{figure}

In Fig.~\ref{fig2} we show our numerical results for the behavior of pion
masses, which are plotted as functions of $eB$. Solid, dashed and
dashed-dotted lines correspond to Sets I, II and III, respectively. In the
case of the $\pi^+$, the curves correspond to the ``magnetic-field dependent
mass'' $E_{\pi^+}$ defined by Eq.~(\ref{epimas}). For comparison we also
show the behavior of $E_{\pi^+}$ in the case of a point-like meson. From the
figure it is seen that, according to the prediction of the model, the
$\pi^+$ structure tends to increasingly enhance the value of $E_{\pi^+}$
when the magnetic field is increased.  The figure also includes the LQCD
results given in Ref.~\cite{Bali:2011qj}, in which values up to $eB\sim
0.4$~GeV$^2$ have been quoted for realistic pion masses using staggered
quarks. It is found that model predictions are in good agreement with LQCD
results for $eB\lesssim 0.15$ GeV$^2$, while they seem to deviate from them
for larger values of the magnetic field. Concerning the $\pi^0$ mass, it is
seen that it shows a slight decrease with $eB$, as previously found e.g.~in
Refs. \cite{Avancini:2015ady,Avancini:2016fgq}. Once again the results are
in general rather independent of the model parametrization.

Besides the mentioned LQCD calculation in Ref.~\cite{Bali:2011qj}, more
recent lattice simulations using Wilson
fermions~\cite{Bali:2017ian,Bali:2017yku} have been carried out, providing
results for $\pi^+$ and $\pi^0$ masses for larger values of $eB$. In these
simulations, however, a heavy pion with $m_\pi(0)=415$~MeV in vacuum has
been considered. In order to compare these results with our predictions we
follow the procedure done in Ref.~\cite{Avancini:2016fgq}, viz.\ we consider
a parameter Set Ib in which $G$ and $\Lambda$ are the same as in Set I,
while $m_0$ is increased so as to obtain $m_\pi(0) = 415$~MeV. Moreover, in
Ref.~\cite{Avancini:2016fgq}, the authors also consider a magnetic field
dependent coupling $G(eB)$ of the form
\begin{equation}
G(eB)= \alpha+\beta\,e^{-\gamma\,(eB)^2}\ ,
\label{geb}
\end{equation}
in order to reproduce LQCD results for the behavior of quark condensates as
well as that of the $\pi^0$ mass.

\begin{figure}[htb]
\centering{}\includegraphics[width=0.7\textwidth]{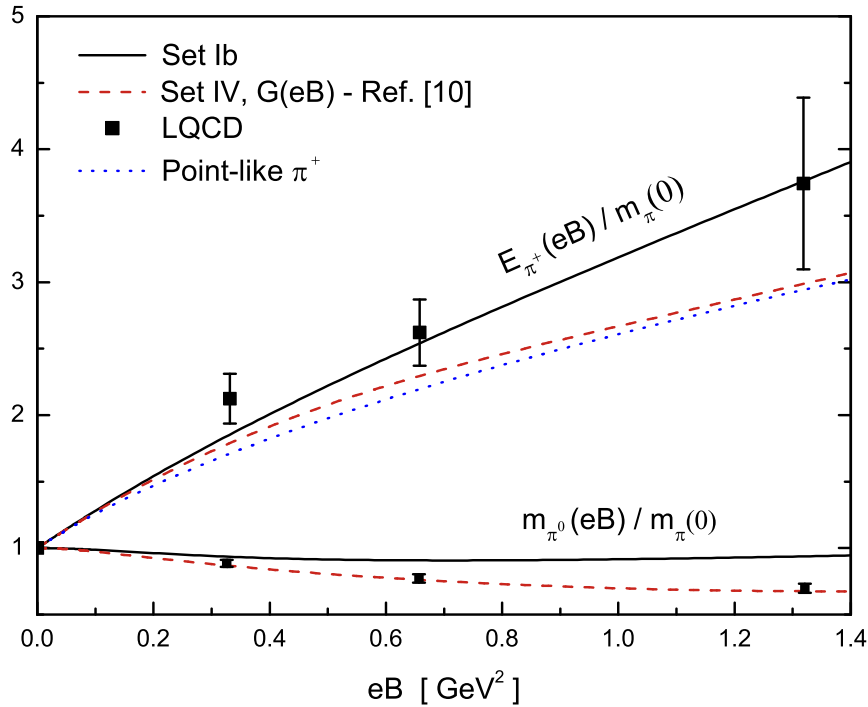} \caption{(Color
online) Normalized neutral pion mass and magnetic field-dependent charged
pion mass as functions of $eB$. Solid lines correspond to the results from
Set Ib, while dashed lines are obtained from Set IV of
Ref.~\cite{Avancini:2016fgq}, considering a magnetic field-dependent
coupling $G(eB)$ as in Eq.~(\ref{geb}). The dotted line shows the behavior
of the normalized magnetic field-dependent mass of a point-like charged
pion, while squares correspond to the results of lattice QCD simulations in
Ref.~\cite{Bali:2017ian}, which consider a $B=0$ pion mass of 415~MeV.}
\label{fig3}
\end{figure}

The curves for the normalized charged pion $B$-dependent mass
$E_{\pi^+}/m_\pi(0)$ and neutral pion mass $m_{\pi^0}/m_\pi(0)$ for Set Ib
are shown in Fig.~\ref{fig3} (solid lines), together with LQCD results
obtained for these quantities after an extrapolation of lattice spacing to
the continuum~\cite{Bali:2017ian}. In addition, we have included in this
figure the results corresponding to the parameter Set IV of
Ref.~\cite{Avancini:2016fgq}, with the $B$-dependent coupling $G(eB)$. It is
seen that for the $\pi^+$ meson the results from Set Ib are consistent with
lattice data, although the errors in the latter are considerably large to be
conclusive (in fact, results obtained considering finite lattice spacings
become closer to those corresponding to a point-like
$\pi^+$~\cite{Bali:2017yku}). On the other hand, in the case of the $\pi^0$
mass, where errors from LQCD are smaller, the curve obtained from Set Ib
appears to be clearly above lattice predictions. Regarding the model
proposed in Ref.~\cite{Avancini:2016fgq}, it is seen that the behavior of
the $B$-dependent mass of the $\pi^+$ is similar to that of a point-like
particle, while (as discussed in Ref.~\cite{Avancini:2016fgq}) the results
for the $\pi^0$ mass are in good agreement with LQCD data. In the case of
that model, it is worth noticing that once $m_0$ is rescaled to get a
phenomenologically acceptable value for the pion mass, the corresponding
parametrization leads to a too low value for the pion decay constant at
$B=0$, namely $f_\pi\simeq 80$ MeV.

\hfill

In conclusion, we have analyzed the effect of an intense external magnetic
field $\vec B$ on $\pi$ meson masses within the two-flavor NJL model. In
particular, we have shown that the Ritus eigenfunction method allows us to
fully take into account the translational-breaking effects introduced into
the calculation of the charged meson masses by the Schwinger phases in the
RPA approach. For the definition of the magnetic-field dependent mass it has
been taken into account that, owing to zero-point motion in the plane
perpendicular to $\vec B$, the charged pion cannot be at rest in the
presence of the magnetic field, even at the lowest Landau level.

In our numerical calculations we have used a model parametrization that
satisfactorily describes not only meson properties in the absence of the
magnetic field but also the behavior of quark condensates as functions of
$B$ obtained in LQCD calculations.  We have found that when the magnetic
field is enhanced, the $\pi^0$ mass shows a slight decrease, while the
magnetic-field dependent mass of the charged pion steadily increases,
remaining always larger than that of a point-like pion. These results are in
agreement with LQCD calculations with realistic pion masses for low values
of $eB$ (say $eB\lesssim 0.15$ GeV$^2$), although there seems to be some
discrepancy as the magnetic field is increased. For larger values of $eB$,
some recent LQCD simulations for $m_{\pi^0}$ and $E_{\pi^+}$ have been
carried out considering unphysically large quark masses. In the case of
$E_{\pi^+}$ the results are consistent with our calculations (with
adequately rescaled parameters), while there is a significant discrepancy in
the case of the $\pi^0$ mass. On the other hand, it is seen that the
agreement for $m_{\pi^0}$ gets improved if, as done in
Ref.~\cite{Avancini:2016fgq}, a magnetic dependent coupling constant $G(eB)$
is introduced. In this sense, we notice that nonlocal NJL-like models, which
naturally predict a magnetic field dependence  of the quark current-current
interaction, have been shown to adequately reproduce the $\pi^0$ mass
behavior~\cite{GomezDumm:2017jij}. A proper analysis of the $\pi^+$ mass in
this framework would be welcome. Concerning the future outlook on this
subject, it is clear that within the NJL model the method used in this work
will allow for a consistent determination of the charged pion decay
constants and the behavior of finite temperature pion properties in the
presence of intense magnetic fields. We expect to report on these topics in
forthcoming publications.

\hfill

NNS is grateful to S.~Avancini for useful correspondence. This work has been
supported in part by CONICET and ANPCyT (Argentina), under grants PIP14-578,
PIP12-449, and PICT14-03-0492, and by the National University of La Plata
(Argentina), Project No.\ X718.


\begin{thebibliography}{199}

\bibitem{Kharzeev:2012ph}
D.~E.~Kharzeev, K.~Landsteiner, A.~Schmitt and H.~U.~Yee,
Lect.\ Notes Phys.\  {\bf 871} (2013) 1.

\bibitem{Andersen:2014xxa}
  J.~O.~Andersen, W.~R.~Naylor and A.~Tranberg,
  Rev.\ Mod.\ Phys.\  {\bf 88} (2016) 025001.

\bibitem{Miransky:2015ava}
  V.~A.~Miransky and I.~A.~Shovkovy,
  Phys.\ Rept.\  {\bf 576} (2015) 1.

\bibitem{HIC}
D.~E.~Kharzeev, L.~D.~McLerran and H.~J.~Warringa, Nucl.\ Phys.\ A {\bf 803} (2008) 227;
V. Skokov, A. Y. Illarionov, and V. Toneev, Int. J. Mod. Phys. A {\bf 24} (2009) 5925;
V. Voronyuk, V. Toneev, W. Cassing, E. Bratkovskaya, V. Konchakovski, and S. Voloshin,
Phys. Rev. C {\bf 83} (2011) 054911.

\bibitem{duncan}
R. C. Duncan and C. Thompson, Astrophys. J. 392 (1992) L9; C.
Kouveliotou et al., Nature {\bf 393} (1998) 235.

\bibitem{Fayazbakhsh:2012vr}
  S.~Fayazbakhsh, S.~Sadeghian and N.~Sadooghi,
  Phys.\ Rev.\ D {\bf 86} (2012) 085042.

\bibitem{Fayazbakhsh:2013cha}
  S.~Fayazbakhsh and N.~Sadooghi,
  Phys.\ Rev.\ D {\bf 88} (2013) 065030.

\bibitem{Avancini:2015ady}
  S.~S.~Avancini, W.~R.~Tavares and M.~B.~Pinto,
  Phys.\ Rev.\ D {\bf 93} (2016) 014010.

\bibitem{Zhang:2016qrl}
  R.~Zhang, W.~j.~Fu and Y.~x.~Liu,
  Eur.\ Phys.\ J.\ C {\bf 76} (2016) 307.

\bibitem{Avancini:2016fgq}
  S.~S.~Avancini, R.~L.~S.~Farias, M.~Benghi Pinto, W.~R.~Tavares and V.~S.~Timóteo,
  Phys.\ Lett.\ B {\bf 767} (2017) 247.

\bibitem{Mao:2017wmq}
  S.~Mao and Y.~Wang,
  Phys.\ Rev.\ D {\bf 96} (2017) 034004.

\bibitem{GomezDumm:2017jij}
  D.~Gomez Dumm, M.~F.~I.~Villafa\~ne and N.~N.~Scoccola,
  Phys.\ Rev.\ D, in press, arXiv:1710.08950 [hep-ph].

\bibitem{Wang:2017vtn}
  Z.~Wang and P.~Zhuang,
  arXiv:1712.00554 [hep-ph].

\bibitem{Liu:2018zag}
  H.~Liu, X.~Wang, L.~Yu and M.~Huang,
  arXiv:1801.02174 [hep-ph].

\bibitem{Andersen:2012zc}
  J.~O.~Andersen,
  JHEP {\bf 1210} (2012) 005.

\bibitem{Agasian:2001ym}
  N.~O.~Agasian and I.~A.~Shushpanov,
  JHEP {\bf 0110} (2001) 006.

\bibitem{Orlovsky:2013wjd}
  V.~D.~Orlovsky and Y.~A.~Simonov,
  JHEP {\bf 1309} (2013) 136.

\bibitem{Andreichikov:2016ayj}
  M.~A.~Andreichikov, B.~O.~Kerbikov, E.~V.~Luschevskaya, Y.~A.~Simonov and O.~E.~Solovjeva,
  JHEP {\bf 1705} (2017) 007.


\bibitem{Bali:2011qj}
  G.~S.~Bali, F.~Bruckmann, G.~Endrodi, Z.~Fodor, S.~D.~Katz, S.~Krieg, A.~Schafer and K.~K.~Szabo,
  JHEP {\bf 1202} (2012) 044.

\bibitem{Hidaka:2012mz}
  Y.~Hidaka and A.~Yamamoto,
  Phys.\ Rev.\ D {\bf 87} (2013) 094502.

\bibitem{Luschevskaya:2014lga}
  E.~V.~Luschevskaya, O.~E.~Solovjeva, O.~A.~Kochetkov and O.~V.~Teryaev,
  Nucl.\ Phys.\ B {\bf 898} (2015) 627.

\bibitem{Bali:2015vua}
  B.~B.~Brandt, G.~Bali, G.~Endrodi and B.~Glaessle,
  PoS LATTICE {\bf 2015} (2016) 265.

\bibitem{Bali:2017ian}
  G.~S.~Bali, B.~B.~Brandt, G.~Endrodi and B.~Glaessle,
  Phys.\ Rev.\ D {\bf 97} (2018) 034505.

\bibitem{Bali:2017yku}
  G.~S.~Bali, B.~B.~Brandt, G.~Endrodi and B.~Glaessle,
  arXiv:1710.01502 [hep-lat].

\bibitem{Vogl:1991qt}
  U.~Vogl and W.~Weise,
  Prog.\ Part.\ Nucl.\ Phys.\  {\bf 27} (1991) 195.

\bibitem{Klevansky:1992qe}
  S.~P.~Klevansky,
  Rev.\ Mod.\ Phys.\  {\bf 64} (1992) 649.

\bibitem{Hatsuda:1994pi}
  T.~Hatsuda and T.~Kunihiro,
  Phys.\ Rep.\  {\bf 247} (1994) 221.

\bibitem{Schwinger:1951nm}
  J.~S.~Schwinger,
  Phys.\ Rev.\  {\bf 82} (1951) 664.

\bibitem{Ritus:1978cj}
  V.~I.~Ritus,
  Sov.\ Phys.\ JETP {\bf 48} (1978) 788.

\bibitem{Klevansky:1989vi}
  S.~P.~Klevansky and R.~H.~Lemmer,
  Phys.\ Rev.\ D {\bf 39} (1989) 3478.

\bibitem{Menezes:2008qt}
  D.~P.~Menezes, M.~Benghi Pinto, S.~S.~Avancini, A.~Perez Martinez and C.~Providencia,
  Phys.\ Rev.\ C {\bf 79} (2009) 035807.

\bibitem{Allen:2015paa}
  P.~G.~Allen, A.~G.~Grunfeld and N.~N.~Scoccola,
  Phys.\ Rev.\ D {\bf 92} (2015) 074041.

\bibitem{Klevansky:1991ey}
  S.~P.~Klevansky, J.~Janicke and R.~H.~Lemmer,
  Phys.\ Rev.\ D {\bf 43} (1991) 3040.

\bibitem{Bali:2012zg}
  G.~S.~Bali, F.~Bruckmann, G.~Endrodi, Z.~Fodor, S.~D.~Katz and A.~Schafer,
  Phys.\ Rev.\ D {\bf 86} (2012) 071502.

\end{thebibliography}
\end{document}